\documentclass[useAMS,usenatbib]{mn2e}
\usepackage{graphicx}

\newcommand{\Mjup}{\mbox{M$_{Jup}$}}
\newcommand{\RHill}{\mbox{R$_{\rm H}$}}
\newcommand{\Msun}{\mbox{M$_{\odot}$}}

\title[A dynamical analysis of the proposed HU Aquarii planetary system]{A dynamical analysis of the proposed HU Aquarii planetary system}

\author[J. Horner, J.P. Marshall, R.A. Wittenmyer \& C.G. Tinney]{J. 
Horner$^{1}$\thanks{E-mail: j.a.horner@unsw.edu.au (JH)}, J. P. 
Marshall$^{2}$, Robert A. Wittenmyer$^{1}$ and C. G. Tinney$^{1}$\\
$^{1}$Department of Astrophysics and Optics, School of 
Physics, University of New South Wales, Sydney 2052, Australia\\ 
$^{2}$Departmento F\'isica Te\'orica, Facultad de Ciencias, Universidad 
Aut\'onoma de Madrid, Cantoblanco, 28049, Madrid, Espa\~na}
\begin{document}

\date{Accepted  Received ; in original form }

\pagerange{\pageref{firstpage}--\pageref{lastpage}} \pubyear{2011}

\maketitle

\label{firstpage}
\vspace{-0.9cm}
\begin{abstract}
It has recently been suggested that the eclipsing polar HU Aquarii is 
host to at least two giant planets. We have performed highly 
detailed dynamical analysis of the orbits of those planets and show that 
the proposed system is highly unstable on timescales of $<$ 
5$\times$10$^{3}$~years. For the coplanar orbits suggested in the discovery letter, we find stable 
orbital solutions for the planetary system only if the outer body moves 
on an orbit that brings it no closer to the host star than $\sim$ 6~AU. The required periastron
distance for the outer planet lies approximately 5 Hill 
radii beyond the orbit of the inner planet, and well beyond the 
1-$\sigma$ error bars placed on the orbit of the outer planet in the 
discovery letter. If the orbits of the proposed planets are 
significantly inclined with respect to one another, the median stability 
increases slightly, but such systems still 
become destabilised on astronomically minute timescales (typically within a few 10$^4$ years). 
Only in the highly improbable scenario where the outer planet follows a retrograde but
coplanar orbit (i.e. inclined by 180$^{\circ}$ to the orbit of the inner planet) is there any significant
region of stability within the original 1-$\sigma$ orbital uncertainties. Our results 
suggest that, if there is a second (and potentially, a third planet) in 
the HU Aquarii system, its orbit is dramatically different to that 
suggested in the discovery paper, and that more observations are 
critically required in order to constrain the nature of the suggested 
orbital bodies.

\end{abstract}


\vspace{-1.0cm}
\section{Introduction}

Precise measurements of timing variations of strictly periodic events 
have been successfully used to infer the existence bodies orbiting around 
distant stars.  Perhaps the best-known examples are the pulsar planets of
\citet{wf92}, a system of three extremely low-mass planets orbiting the 
millisecond pulsar PSR1257+12. Pulsating sub-dwarf B stars have also 
been found to host planets, e.g.~V391 Peg \citep{silvotti07} and HW Vir 
\citep{lee09}.  This same technique has recently been applied to 
eclipsing polars, using the egress of the small, bright accretion spot 
as a precise ``clock.'' Unseen orbiting bodies can cause small shifts in the timing of eclipses 
(ranging from $\sim$10~sec to a few minutes) due to the 
light-travel time differences imposed by the gravitational influence of 
the orbiting bodies on the system barycentre. \citet{qian10} discovered 
the first such planet orbiting the eclipsing polar DP Leo by combining 
their observed eclipse timings with a long-term data set from 
\citet{schwope02}; the combined data set revealed a sinusoidal variation 
indicative of a 6.3\Mjup~planet with a period of 23.8~years.  The 
timing method has shown that planets can orbit stars which are wildly 
different from the main-sequence solar-type stars most commonly targeted 
by Doppler and transit planet search programs.

In a recent letter, \citet{Qian2011} announced the discovery 
of two (and potentially more) 
giant planets orbiting the 
eclipsing polar HU Aquarii (hereafter HU Aqr). The authors provided fits to the orbits of 
those planets, placing them at orbital radii of 3.6 and 5.4~AU, from the 
system barycentre, and ascribed minimum masses of 5.9 and 4.5~\Mjup (respectively)
to the two bodies.  In this Letter, we perform a detailed 
dynamical analysis of the HU Aqr planetary system in order to assess the 
stability of the proposed planet candidates.

\vspace{-0.7cm}
\section{The HU Aquarii Planetary System}

In their study of the eclipsing polar system HU Aqr, 
\citet{Qian2011} consider the temporal variation in the observed eclipse 
timings (``O'') as compared to predicted timings (``C'') that would be expected from a 
linear ephemeris. By plotting a simple 
$O-C$ diagram, they show that the $O-C$ residuals for HU 
Aqr contain two cyclical signals superposed on a longer-period 
curvature.  Each signal can be modelled as a Keplerian orbit to 
determine the planetary parameters.  We reproduce the parameter 
estimates of \citet{Qian2011} in Table~\ref{planetparams}. As 
for planets detected with by radial-velocity method, only the radial 
component of the planet's influence on the host star is detectable.  
Here, only the line-of-sight light-travel time differences are observed, 
so the mass estimates for the HU Aqr planets are given as minimum 
values. \citet{Qian2011} note that the HU Aqr system inclination is 
$85^{\circ}$, so, if the planets orbit in the same plane as the stars, 
their true masses would only be 0.4\% larger than the minimum values 
given in Table~\ref{planetparams}. 

\begin{table}
  \centering
  \begin{tabular}{lll}
  \hline
Parameter & HU Aqr (AB)b& HU Aqr (AB)c\\
 \hline
Eccentricity                     & 0.0   & 0.51$\pm$0.15 \\
Orbital Period (yrs)             & 6.54$\pm$0.01 & 11.96$\pm$1.41\\
Orbital Radius (AU)              & 3.6$\pm$0.8  & 5.4$\pm$0.9\\
Minimum Mass (\Mjup) & 5.9$\pm$0.6  & 4.5$\pm$0.5\\
 \hline
 \end{tabular}
  \caption{The orbits of the HU Aqr exoplanets \citep{Qian2011}.}
\label{planetparams}
\end{table}

\vspace{-0.6cm}
\section{A dynamical search for stable orbits}

In order to examine the potential dynamical stability of the two planets 
suggested for the HU Aqr system, we performed a large number of 
detailed dynamical simulations using the \textit{Hybrid} integrator within 
the $N$-body dynamical package \textit{Mercury} \citep{Chambers99}. Following 
the strategy employed to analyse the stability of the 
HR~8799 system \citep{Marshall2010}, we held the orbit of the inner 
planet constant (with $a = 3.6~AU$ and $e = 0.0$), and varied the 
orbital elements of the outer planet across a range corresponding to $\pm$3 times the discovery 
letter's quoted uncertainties in the semi-major axis, $a$, and eccentricity, $e$.
We initially considered the scenario described in \citet{Qian2011}, 
where the planets are considered to be co-planar. In other words, we set 
the orbital inclinations, $i$, of the two planets to be 0$^{\circ}$ at the 
start of our integrations.  We treat the central stars, a 0.88~\Msun~white dwarf and a 
0.2~\Msun~secondary, as a single point mass.  Since 
the stars orbit each other with a period of only 2.08 hours (i.e.~with a 
separation of 0.004~AU), and the bodies of interest are believed to orbit at distances 
of 3.6 and 5.4~AU, this treatment is dynamically justified. We give each planet the minimum 
mass estimated in \citet{Qian2011}. We note in passing that, 
if the masses of the proposed planets are significantly greater than those detailed
in \citet{Qian2011}, then this could only have a deleterious effect on the stability of their proposed orbits.

Fixing the orbit of the inner planet, we simulated a total of 
9261 planetary systems. In each simulated system, the inner planet 
began on the same orbit, but the orbital elements of the outer planet 
were chosen such that each simulation sampled a unique set of possible parameters. We 
distributed the orbital elements of the outer planet such that we tested 
21 values of semi-major axis, spread evenly across $\pm$3-$\sigma$ from the value ($a=5.4$~AU) given in \citet{Qian2011}. For 
each value of semi-major axis, we tested 21 values of orbital 
eccentricity, again spread evenly across $\pm$3-$\sigma$ from the 
value ($e=0.51$), given in that work. Finally, at each of these 441 
 $(a,e)$ locations we carried out 21 tests, with the initial location of the planet distributed across a 
range of $\pm$3-$\sigma$ from the nominal mean anomaly of the outer planet 
(calculated from Fig. 2 of \citet{Qian2011}).  These 9261 unique models, 
based on the HU Aqr system parameters, were integrated using the 
\textit{Hybrid} integrator within \textit{Mercury} for a period of 100~Myr, 
following the evolution and final fates of the two postulated planets in the system. 
A planet was deemed ejected from the system upon 
reaching a distance of 1000~AU from the barycentre, and all mutual collisions were recorded. 
This yielded a lifetime from each 
individual integration in the range 0 - 100~Myr, defined as the time 
until one or other of the planets was removed from the system through 
either collision or ejection.

To explore the 1-$\sigma$ parameter range in greater detail, we launched a
second suite of integrations, again 
using 9261 test systems. The set-up was performed exactly as 
described above, except that the orbital parameters of the outer 
planet were varied within a 1-$\sigma$ range, rather than the 3-$\sigma$ 
distribution carried out previously.

These two suites of integrations yielded 17493 
distinct tests of the stability of the HU Aqr system, with 9261 of 
these performed in the central $\pm$1-$\sigma$ of the element space 
described in \citet{Qian2011}, and the other 8232 distributed in the 
range 1-$\sigma$ to 3-$\sigma$. The results of these integrations are shown in Fig~\ref{coplanar}.

Once these simulations had been completed, we carried out equivalent 
suites for scenarios where the outer planet was moving on an orbit 
inclined to that of the inner planet. Following the procedure 
detailed above, we tested systems in which the outer planet's orbit 
was inclined by 5$^{\circ}$, 15$^{\circ}$ and 45$^{\circ}$ with respect to the inner's 
orbit, in order to examine the influence of mutual orbital inclinations 
on the stability of the system. We then considered further scenarios in 
which the outer planet was moving in a retrograde sense, with 
respect to the inner body, with inclinations of 135$^{\circ}$ and 180$^{\circ}$. 
%TEXT ADDED BY JONTI IN RESPONSE TO REFEREE'S MINOR POINT #2
For simplicity, we kept the masses of the two planets constant through these 
runs at the lowest values suggested in \citet{Qian2011}. Although it is true that 
significantly inclined orbits for the planets would result in larger real masses for them,
we note that a significant mutual inclination between the planets does not necessarily 
mean that it is the outermost planet that is inclined to the line of sight, while the innermost is 
in that plane. Rather than attempt to shift the planetary mass by the free parameter of potential
inclination, we instead took the lowest masses possible. We remind the reader that this essentially
means that we have allowed the planetary system the best possible chance of being stable - as the
mass of the planets increases, so does their gravitational reach, increasing the strength of any mutual
interactions.
%END OF ADDED TEXT

\vspace{-0.6cm}
\section{Results} 

\begin{figure}
\includegraphics[scale=0.40,angle=90]{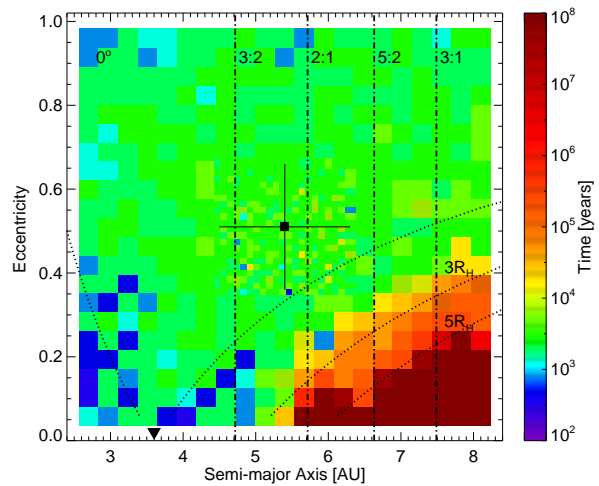}
\caption{Lifetime stability plot of the HU Aqr system using 
the parameters of \citet{Qian2011} with the planets on coplanar orbits. 
The simulations detailed cover the 3-$\sigma$ parameter space in a grid 
of 21$\times$21 points. The central 1-$\sigma$ parameter space is covered by an 
additional and denser set of 21$\times$21 points. Each grid point is the median lifetime of 21 simulated HU Aqr 
systems with the outer planet's initial $a$ and $e$. The location of the inner planet
is denoted by the point of the solid triangle on the x-axis, while the nominal orbit of the outer planet is marked 
by the filled square. Solid lines show the extent of the 1-$\sigma$ errors
in $a$ and $e$ suggested by \citet{Qian2011}. The vertical dot-dashed lines show the location of 
the strongest MMRs in relation to the orbit of the inner planet. The two dotted
lines radiating from the location of the inner planet connect all orbits of that have 
either their periastron (outward curving line) or apastron (inward curving line) at 
the location of the planet. As such, all orbits in the region bounded by these lines 
cross the orbit of the inner planet. The dotted lines labelled $3~R_{H}$ and $5~R_{H}$ 
connect all orbits that pass periastron at a distance of three or five Hill radii beyond the orbit of the inner planet.
It can readily be seen that the planetary system 
shows extreme instability, aside from in a small region of $a$-$e$ space 
to the lower right hand side of the figure, corresponding to orbits of 
the outer planet that remain at a barycentric distance beyond $\sim$6~AU.}
\label{coplanar}
\end{figure}

The results of our $i=$0$^{\circ}$ (i.e. coplanar) dynamical integrations are shown in Fig~\ref{coplanar}. 
Each colour box in that figure shows the median lifetime obtained from 21 independent integrations performed with 
the outer planet placed on an orbit with that particular combination of $a$ and
$e$. The location of the inner planet is marked with a hollow circle, whilst the location of the nominal orbit for the outer planet
given in \citet{Qian2011} is shown by a hollow square, with the 1-$\sigma$ error bars given in that work denoted by the solid lines stretching from that square.
The vertical dot-dash lines show the locations of the strongest mean-motion resonances (hereafter MMR) 
with the orbit of the inner-most planet. The curved dotted lines that meet at the location of the inner planet join all 
orbits whose periastron (outward curving line) or apastron (inward curving line) lie at a distance equal to the orbital 
radius of the inner planet. Every point in the region above these lines denotes orbits for the outer 
planet that cross that of the inner. The two dotted lines labelled 3$R_{Hill}$ and 5$R_{Hill}$ connect orbits which pass 
periastron at a distance equal to the orbital radius of the inner planet plus 3 and 5 times that planet's Hill radius (\RHill), respectively, where \RHill~is defined as

\begin{equation}
R_{H} = a_{p} \Big({M_{p} \over {3M_{s}}}\Big)^{1/3}
\end{equation}

The Hill radius, \RHill, is commonly used in studies of orbital dynamics as a proxy for the 
dynamical ``reach'' of a given body \citep{Horner2003, Horner2004a, 
Horner2004b}. Close encounters between two massive bodies are typically 
defined as those that occur at a distance closer than 3 \RHill, 
although some particularly conservative studies consider 5 \RHill~a 
sufficiently close approach to be labeled as such. These lines 
therefore show the limits at which the outer planet approaches the 
inner within the prescribed number of \RHill~-- those orbits 
outside the lines are too widely separated at the start of the 
integrations to undergo close encounters, while all those within the 
region bounded by the periastron and apastron lines for the orbital 
radius of the inner planet are orbits which cross that of that planet.

It is immediately apparent in Fig~\ref{coplanar} that the great 
majority of possible orbits suggested in \citet{Qian2011} are extremely unstable, with just a small 
region at low eccentricities and high semi-major axes (i.e. below the 3 \RHill~line in the bottom right-hand corner) 
showing any significant long term stability. This is not a surprising result - the 
two planets in question have particularly large masses, and therefore 
have very large dynamical reaches - and so must be widely separated in 
order that they do not strongly perturb one another. Indeed, it is clear from 
that figure that the regions of stability and instability for the HU Aqr system are a strong 
function of the peri- and apastron distances of the initial orbits of 
the outer planet. The only orbits that display relatively strong 
stability all have periastron at distances greater than 5 \RHill~beyond 
the orbit of the inner planet. Any closer, and the encounters 
between the planets are strongly disruptive. Orbits of the outer planet which approach the inner planet to a distance
between 3 and 5 \RHill~are clearly more stable than those which come closer to that planet, but still display
significant instability on astronomically short timescales.

\begin{figure*}
\centering
\includegraphics[scale=0.62]{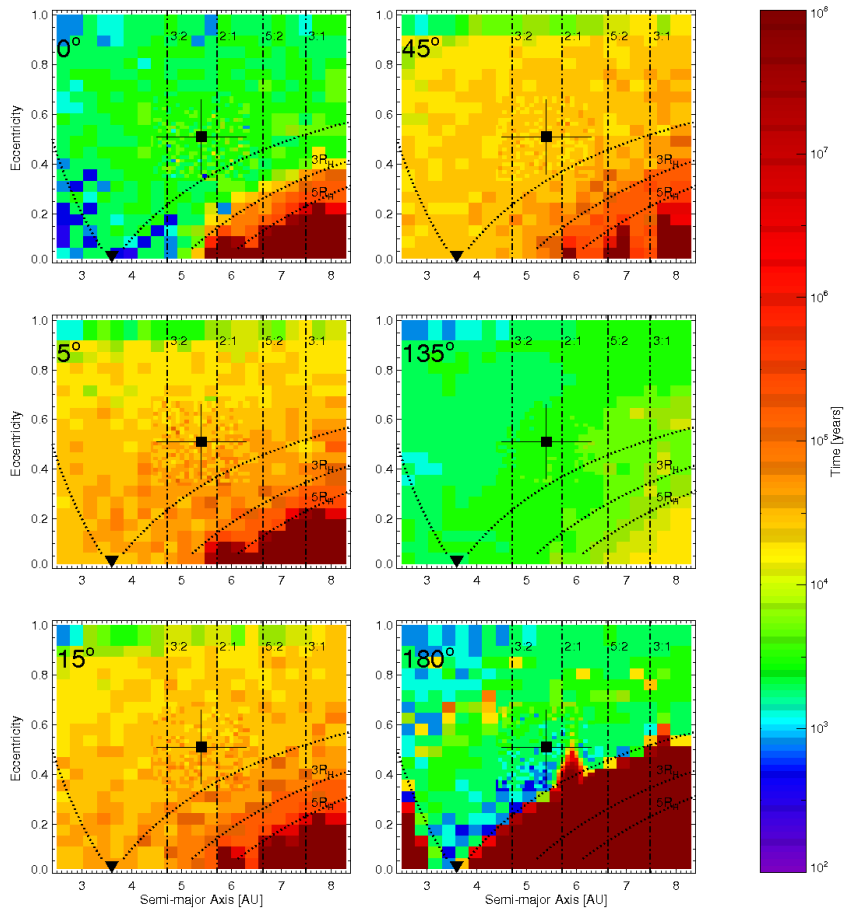}
\caption{Lifetime $a$-$e$ stability plots of the HU Aqr system using 
the parameters of \citet{Qian2011}. The top-left panel reproduces the results
shown in Fig~\ref{coplanar}, detailing the scenario in which both planets are considered
to be moving on co-planar, prograde orbits. The panel to the middle-left shows the
results for the scenario where the outer planet has an orbit inclined by 5$^{\circ}$ to that
of the inner planet, with the lower-left, top-right, middle-right and lower-right panels showing
the results for initial orbital inclinations of 15$^{\circ}$, 45$^{\circ}$, 135$^{\circ}$ and 180$^{\circ}$, respectively. The various features
plotted on each panel are the same as those shown in Fig~\ref{coplanar}.}
\label{sixplot}
\end{figure*}

However, what of mutual inclinations between the two planets? Could the 
system be stabilised by the planets moving on orbits that are 
significantly inclined with respect to one another? This question can be 
answered by examination of Fig~\ref{sixplot}, which presents the results for all six 
scenarios considered in this work. 

As the orbital inclination of the outer planet is increased to 5$^{\circ}$ and then 15$^{\circ}$, the overall stability of the system appears to increase somewhat, with the yellow and orange colours that denote moderate stability spreading across the entire plot. However, the effect of increased inclination on the region of greatest stability, those orbits with median lifetimes greater than one million years, is negligible. Indeed, by the time the inclination is increased to 15$^{\circ}$, that region appears to shrink somewhat, with particular destabilisation occurring just inwards of the location of the 5:2 MMR with the orbit of the inner planet. Once the orbital inclination of the outer planet is increased to 45$^{\circ}$ (top right hand panel), this effect becomes far more pronounced, with only small regions of stability remaining in low-eccentricity orbits around the 2:1 MMR, around the 5:2 MMR, and a larger region beyond the location of the 3:1 MMR. Indeed, far from enha
 ncing the stability of the planetary system, as might be expected, the 3:1 MMR acts to destabilise the orbits of the planets in this more highly inclined scenario. 

The central panel on the right hand side of Fig~\ref{sixplot} shows the scenario for which the orbit of the outer planet is inclined by 135$^{\circ}$ to that of the inner. That scenario displays a remarkable lack of stability across the entire 3-$\sigma$ region of plausible orbits for the outer planet. 

The only scenario in which orbits within the 1-$\sigma$ uncertainty range for the outer planet display strong stability is shown in the lower-right hand 
panel of Fig.~\ref{sixplot} ($i=$180$^{\circ}$), when the orbit is both retrograde and coplanar. In that scenario, almost all orbits for the outer planet 
not crossing that of the inner are dynamically stable on long timescales. In that extreme scenario, even some configurations in which the orbits of the 
two planets cross show significant stability. Those scenarios are the only ones of our entire suite of almost 105,000 test integrations in which orbits of 
the HU Aqr planets situated within the 1-$\sigma$ error bounds stipulated in \citet{Qian2011} can display any long-term stability.

\vspace{-0.6cm}
\section{Conclusions and Discussion}

We have investigated the stability of the recently discovered planetary system around HU Aqr across the $\pm$3-$\sigma$ uncertainty ranges
for the orbit of the outer planet. In the simplest, coplanar, case, we find that the system is unstable on 
timescales $<~5\times10^{3}$ years, except for scenarios in which the 
mutual separation between the two planets was greater than 5~\RHill~at 
the outer planet's periastron. When the orbit of the outer planet is 
set such that it reaches periastron between 3~\RHill~and 5~\RHill~beyond the orbit of the inner planet,
the orbital stability is somewhat greater than scenarios where the minimum separation between the two planets
is smaller then 3~\RHill, but the planetary system still falls apart on timescales too short to inspire any confidence that 
the tested orbits represent the true state of the HU Aqr system. We see a steady increase in the 
median lifetime of the system across much of the tested phase-space as the mutual inclination of the orbits 
climbs from the coplanar case, through 5$^{\circ}$ to 15$^{\circ}$ to 45$^{\circ}$, as expected from the resulting
reduction in the amount of time the two planets spend in close proximity. However, even in the 
most extreme prograde case investigated ($i = 45^{\circ}$), the lifetime of systems within the 
1-$\sigma$ parameter space did not exceed $10^{5}~$years. This is much 
shorter than the expected system age, suggesting that we have either caught the 
system during a period of significant dynamical instability in which the planets are 
undergoing a rearrangement/ejection, or that the orbital solution described in \citet{Qian2011} 
does not represent the true state of the system. Interestingly, although an increase in the mutual inclination of the 
planetary orbits causes those orbits in the unstable region to become slightly more stable, it also results in a reduction 
in the stable region at distances greater than 5~\RHill~from the inner planet.

If the orbit of the outer planet is retrograde compared to that 
of the inner planet, but the orbits remain coplanar (i.e. the mutual inclination of the two orbits 
is 180$^{\circ}$, then we find that the system could be stable across a wide range of parameter space, including some
scenarios within the 1-$\sigma$ errors quoted for the orbit of the outer planet. However, it seems difficult to comprehend
how the proposed planets could have evolved into such orbits. By contrast, 
if the orbits of the two planets have mutual inclinations of 135$^{\circ}$, then 
we find that no region of the tested $a$-$e$ phase space is stable on timescales greater than $\sim 10^{4}$~years.

The results of our dynamical simulations raise a number of interesting 
possibilities. Assuming that the detection of the planets by 
\citet{Qian2011} is robust, and that the orbits of the planets at the 
current epoch are exactly as described in that work, the planetary 
system must be going through a dramatic period of dynamical instability, 
which will in short order result in the loss of one (or both) of the 
planets therein. However, the incredibly short lifetimes we find for the 
proposed planets suggest that, statistically, this is unlikely 
(the odds of observing a planetary system during the last few thousand 
years of a multi-billion-year lifetime seem remarkably small). On the 
other hand, if we assume that the detection of the planets is robust, 
but the orbital parameters given are not a good measure of the true 
state of the system, we suggest that it is most likely that the 
outer planet is moving on a low-eccentricity orbit far from the 
central bodies. Such an orbit would allow that planet to remain 
sufficiently far from the inner planet to be dynamically stable on 
multi-million year timescales, and therefore seems a much more 
reasonable solution for the dynamics of the system. If the two planets 
detected by \citet{Qian2011} have significant mutual orbital 
inclinations, then we find that a slightly wider range of orbits are 
possible for the outer planet to display moderate stability, but that at the same time the region
of greatest stability for that planet's orbit actually reduces in size. In any case, in such scenarios
the stable regions remain relatively restrictive both in terms of orbital eccentricity and semi-major axis - 
keeping the planet well beyond the inner's sphere of influence.

The only way in which the orbits of the two planets can be induced to show significant stability is to 
consider a scenario in which they are coplanar, but with the orbit of the outer planet being retrograde with 
respect to that of the inner. In such a scenario, a wide area of the studied $a$-$e$ phase space becomes 
dynamically stable, including some solutions within the 1-$\sigma$ uncertainties detailed by \citet{Qian2011}.
Whilst this might appear promising, we note that it is hard to envision a way in which the planets could evolve into 
such an unusual configuration without significantly destabilising one another's orbits.

When one examines the residuals from the O-C diagram in the 
\citet{Qian2011} work, a first glance suggests that a low-eccentricity 
orbit for the outer planet is incompatible with the observed data. However, it is 
possible that the removal of the long-term quadratic trend by \citet{Qian2011} could have 
resulted in an artificially enhanced eccentricity for the outer planet, reducing its stability. 
On the other hand, detailed simulations of Doppler velocity data by 
\citet{eccentric} demonstrated that two planets in circular orbits can 
mimic the signal of a single planet on an eccentric orbit, so long as those 
planets move on mutually resonant orbits. This possibility was 
recently explored by \citet{tinney11} for the case of HD~38283b, an 
eccentric planet in a one-year orbit. If, rather than a highly 
eccentric second planet, we have a scenario where the system contains at 
least three massive planets, each moving on low-eccentricity orbits, it 
might be possible to explain the suggested shape of the O-C diagram in 
\citet{Qian2011} whilst placing the planets on orbits that are 
dynamically stable. Indeed, \citet{Qian2011} suggest that there might be 
a third massive body at a large barycentric distance, based on the 
archaic Titius-Bode law. While the use of that law is not generally 
encouraged as a predictive tool in exoplanetary science, the presence of 
a more distant third body would allow orbital fits with the second 
planet moving on a low-eccentricity orbit. Although we have not 
dynamically simulated such a speculative scenario, it seems reasonable 
to assume that if a third planet were located at an orbital radius at 
least 5 \RHill~beyond that of the second planet, the system could display 
long term dynamical stability without violating the observed variations 
in the egress of the accretion hotspot from eclipse by the stellar 
secondary body.

A more detailed statistical analysis of this 
highly fascinating exoplanetary system is clearly necessary in order to disentangle the true nature 
of the proposed planetary system. Such work will doubtless throw fresh light on one of
the most peculiar planetary systems detected to date.

\vspace{-0.6cm}
\section*{Acknowledgments}
JH gratefully acknowledges the financial support of the Australian 
government through ARC Grant DP0774000. RW is supported by a UNSW 
Vice-Chancellor's Fellowship. JPM is partly supported by Spanish grant 
AYA 2008/01727, and gratefully acknowledges Maria Cunningham for funding 
his collaborative visit to UNSW. We also wish to thank the anonymous referee for their 
swift and very helpful feedback.

\label{lastpage}

\end{document}